\documentclass[twocol]{ametsocV5}

\usepackage{amsmath,amsfonts,amssymb,bm}
\usepackage{mathptmx}%{times}
\usepackage{newtxtext}
\usepackage{newtxmath}

\title{A simple model for predicting the hurricane radius of maximum wind from outer size \\
{\color{red}NOT PUBLISHED. Submitted for peer review}}

    \authors{Daniel R. Chavas\correspondingauthor{Daniel R. Chavas, drchavas@gmail.com}}
     \affiliation{Purdue University, Department of Earth, Atmospheric, and Planetary Sciences, 
     West Lafayette, Indiana}

    \extraauthor{John A. Knaff}
    \extraaffil{NOAA Center for Satellite Applications and Research, Fort Collins, Colorado}

\abstract{The radius of maximum wind ($R_{max}$) in a hurricane governs the footprint of hazards, particularly damaging wind and rainfall. However, $R_{max}$ is noisy to observe directly and is poorly resolved in reanalyses and climate models. In contrast, outer wind radii are much less sensitive to such issues. Here we present a simple empirical model for predicting $R_{max}$ from the radius of 34-kt wind ($R_{17.5ms}$) that only requires as input quantities that are routinely estimated operationally: maximum wind speed, $R_{17.5ms}$, and latitude. The form of the empirical model takes advantage of our physical understanding of hurricane radial structure and is trained on the Extended Best Track database from the North Atlantic; results are similar for the TC-OBS database. The physics reduces the relationship between the two radii to a dependence on two physical parameters, while the observational data enables an optimal estimate of the quantitative dependence on those parameters. The model performs substantially better than existing operational methods for estimating $R_{max}$. The model reproduces the observed statistical increase in $R_{max}$ with latitude and demonstrates that this increase is driven by the increase in $R_{17.5ms}$ with latitude. Overall, the model offers a simple and fast first-order prediction of $R_{max}$ that can be used operationally and in risk models.}

\begin{document}

%% Necessary!
\maketitle

%%%%%%%%%%%%%%%%%%%%%%%%%%%%%%%%%%%%%%%%%%%%%%%%%%%%%%%%%%%%%%%%%%%%%
% SIGNIFICANCE STATEMENT/CAPSULE SUMMARY
%%%%%%%%%%%%%%%%%%%%%%%%%%%%%%%%%%%%%%%%%%%%%%%%%%%%%%%%%%%%%%%%%%%%%
%
% If you are including an optional significance statement for a journal article or a required capsule summary for BAMS 
% (see www.ametsoc.org/ams/index.cfm/publications/authors/journal-and-bams-authors/formatting-and-manuscript-components for details), 
% please apply the necessary command as shown below:
%
\statement
If we can better predict the area of strong winds in a hurricane, we can better prepare for its potential impacts. This work develops a simple model to predict the radius where the strongest winds in a hurricane are located. The model is simple and fast and more accurate than existing models, and it also helps us to understand what causes this radius to vary in time, from storm to storm, and with changes in latitude. It can be used in both operational forecasting and models of hurricane hazard risk.
%
% \capsule
% Capsule summary here.

%%%%%%%%%%%%%%%%%%%%%%%%%%%%%%%%%%%%%%%%%%%%%%%%%%%%%%%%%%%%%%%%%%%%%
% MAIN BODY OF PAPER
%%%%%%%%%%%%%%%%%%%%%%%%%%%%%%%%%%%%%%%%%%%%%%%%%%%%%%%%%%%%%%%%%%%%%
%

\section{Introduction}

The radius of maximum wind ($R_{max}$) in a hurricane principally determines the location and areal extent of storm hazards, including extreme wind and rainfall \citep{Lonfat_Rogers_Marchok_Marks_2007, Lu_etal_2018, Xi_Lin_Smith_2020} and storm surge \citep{Penny_etal_2021,Irish_Resio_Ratcliff_2008,Irish_Resio_2010}. Hence, a simple first-order prediction of $R_{max}$ has significant value for both operational forecasting and risk assessment. However, $R_{max}$ is noisy and difficult to observe directly in real storms due to the turbulent nature of the moist convective inner core boundary layer \citep{Shea_Gray_1973,Kossin_etal_2007,Sitkowski_Kossin_Rozoff_2011,Kepert_2017,Stern_etal_2020}. $R_{max}$ is also poorly resolved by numerical weather prediction models, especially climate models, as low horizontal resolution acts to smooth the inner core structure radially outwards \citep{Reed_Jablonowski_2011,Gentry_Lackmann_2010,Rotunno_Bryan_2012}. In contrast, the outer storm circulation is relatively quiescent and hence less variable in space and time \citep{Frank_1977,Cocks_Gray_2002,Chavas_Lin_2016}. As a result, a measure of the size of the broad outer circulation (``outer size"), such as the radius of 34-kt wind, is much easier to resolve, less sensitive to turbulence, and is predictable operationally \citep{Knaff_Sampson_2015}. Could we use information about outer size to help predict $R_{max}$?

To do so, we need a means of linking outer size to $R_{max}$: storm structure. Fundamentally, $R_{max}$ is strongly dependent on the outer circulation because the source of angular momentum at $R_{max}$ is inward advection from larger radii \citep{Palmen_Riehl_1957}. Moreover, because surface friction removes angular momentum from air parcels as they spiral inwards towards $R_{max}$, angular momentum must gradually decrease moving inwards towards $R_{max}$. \citet{Chavas_Lin_Emanuel_2015} developed a physical model for the complete radial structure of the low-level angular momentum distribution and, in turn, the cyclonic wind field (we refer to this model hereafter as "C15"). \citet{Chavas_Lin_2016} demonstrated that this model predicts the qualitative dependence of $R_{max}$ on key parameters that is found in observations: $R_{max}$ tends to be larger for smaller $V_{max}$, larger outer size, and higher storm center latitude. This result complements preceding empirical work showing similar dependencies on intensity and latitude \citep{Kossin_etal_2007,Knaff_etal_2015}. Hence, the C15 model appears to capture the fundamental first-order physics linking outer size to $R_{max}$. The structural model also provides the foundation for understanding the relationship between the hurricane minimum central pressure and maximum wind speed \citep{Knaff_Zehr_2007, Courtney_Knaff_2009, Chavas_Reed_Knaff_2017}. Moreover, this approach explicitly identifies a small number of physical parameters that govern how $R_{max}$ depends on $V_{max}$, latitude, and outer size.

In principle, this wind structure model could be used to directly predict $R_{max}$ from outer size \citep[e.g.][]{Davis_2018}. However, this physical model is idealized and thus has biases relative to observations in representing the relationships between radii \citep{Chavas_Lin_Emanuel_2015, Chavas_Lin_2016}. In contrast, one may predict $R_{max}$ using purely empirical regression models based on observed quantities such as storm intensity or satellite-derived quantities \citep{Mueller_etal_2006,Willoughby_Darling_Rahn_2006,Kossin_etal_2007,Vickery_Wadhera_2008,Knaff_etal_2015}. However, this approach ignores valuable information encoded in the physics of storm structure described above that can help constrain the prediction of $R_{max}$.

An ideal alternative is a hybrid model that takes advantage of both physical and observational information simultaneously. One way to do this is to exploit the theoretical basis for the C15 structure model, rather than the model itself, in the design of an empirical regression model fit to observations. This approach would use the physics to dictate model predictors and predictands but then use observations to define their true relationship in nature. Moreover, such a model may potentially be applied to predict how $R_{max}$ may change in a future climate state, where no observational data are available.

Given the above knowledge gap, the principal objectives of this work are:
\begin{enumerate}
    \item To develop a simple predictive model for $R_{max}$ from outer size (here the radius of 34-kt wind) that exploits the benefits of both observations and physical theory,
    \item To estimate model parameters from historical hurricane data in the North Atlantic basin,
    \item To compare model performance against existing predictive models for $R_{max}$.
\end{enumerate}
In pursuing these objectives, we aim to predict but also to explain \citep{Shmueli_2010}: our primary goal is a model to predict $R_{max}$ that is useful for everyday applications, but in the process we hope the model will also provide insight into the underlying physics linking $R_{max}$ to outer size.

Section \ref{sec:methods} describes the methodology. Section \ref{sec:results} presents the results of our model and comparison with other existing predictive models. Finally, Section \ref{sec:conclusions} summarizes our findings and discusses avenues for future work.

\section{Methodology}\label{sec:methods}

Our goal is to develop a model to predict $R_{max}$ from outer size that combines information linking the two quantities from both the underlying physics of hurricane structure and observations. The physics will simplify our model to depend on only two predictive physical parameters, and then observational data will be used to fit the coefficients of those predictors in a simple empirical model to ensure they most closely match that found in real storms. We begin with a conceptual overview of our model design. We then describe the theory for the physical component of the model. Finally, we describe the datasets and regression methodology used to define the final predictive model.

For our purposes, we focus in this paper exclusively on the radius of 34kt wind ($17.5 \; ms^{-1}$; hereafter $R_{17.5ms}$) as our measure of outer size, as this is the outermost wind radius that is commonly estimated in operations. $R_{17.5ms}$ is also typically located in the outer circulation where convection is minimal (with the exception of very weak storms) and thus it tends to covary minimally with $V_{max}$ \citep{Merrill_1984,Chavas_Lin_2016}. However, this approach can in principle be applied to any wind radius if given appropriate data.

% To ensure unit consistency and avoid confusion when doing calculations, we write $34kt$ in MKS units as $17.5ms$ throughout this work.

\subsection{Conceptual overview}

\begin{figure*}[h]
\centerline{\includegraphics[width=\textwidth]{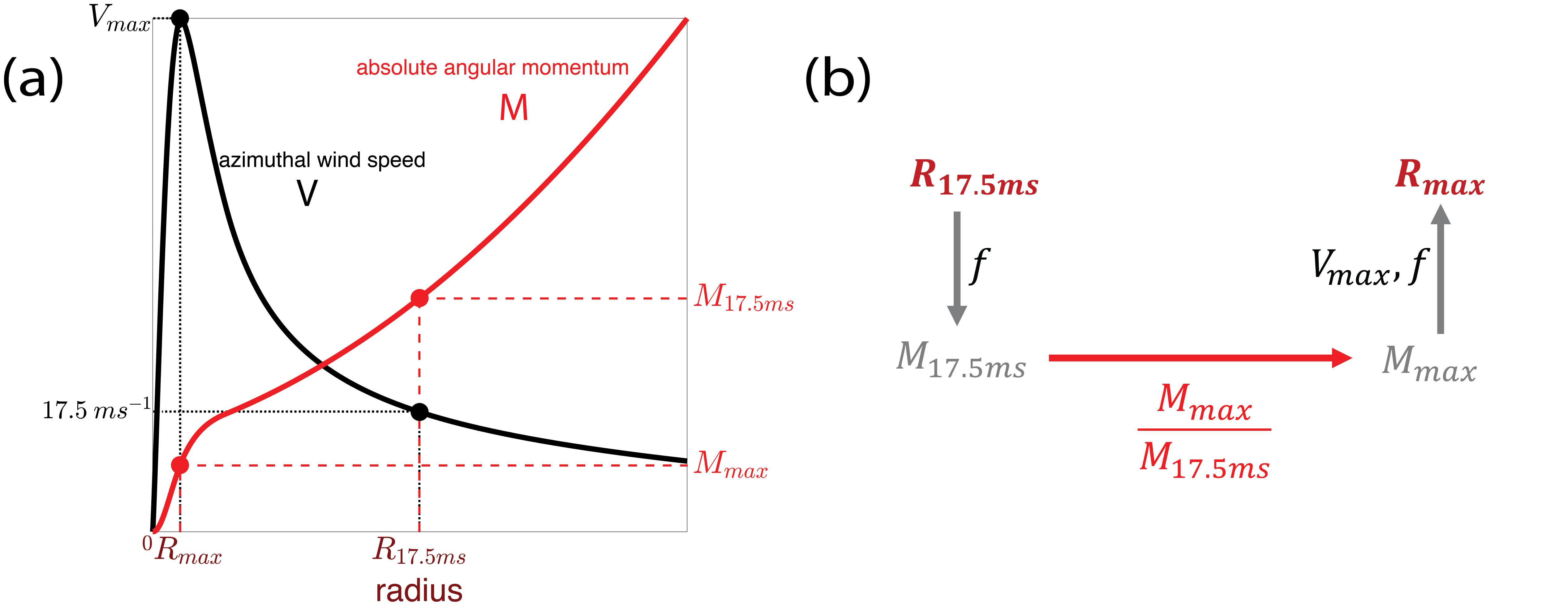}}
\caption{(a) Example radial structure of azimuthal wind speed (black) and absolute angular momentum (red), with relevant radii denoted. (b) Conceptual framework of our model for predicting $R_{max}$ from $R_{17.5ms}$ via eqs. \eqref{eq:Mouter}-\eqref{eq:Rmax}.}
\label{fig:conceptual}
\end{figure*}

A conceptual diagram of our model is shown in Figure \ref{fig:conceptual}. Hurricanes are overturning circulations in which boundary layer air parcels at larger radii flow radially inwards towards the center in the presence of background rotation \citep{Riehl_1950,Wing_Camargo_Sobel_2016}. The absolute angular momentum (hereafter simply ``angular momentum'') of an air parcel is given by
\begin{equation}\label{eq:M}
M = rV + \frac{1}{2}fr^2
\end{equation}
where $r$ is radius, $V$ is the tangential wind speed, $f=2\Omega sin\phi$ is the Coriolis parameter, $\Omega=7.292*10^{-5} \; s^{-1}$ is the Earth's rotation rate, and $\phi$ is the latitude of the storm center. The first term is the relative angular momentum associated with the storm circulation. The second term is the planetary angular momentum associated with the projection of the Earth's rotation onto the hurricane's axis of rotation (vertical) at the latitude of the storm center\footnote{The planetary angular momentum term can be written as $(\Omega sin\phi r)*r$, where the term in parentheses is the tangential velocity of the Earth's rotation projected onto the local vertical. Thus, even at the outer edge of the storm where the circulation vanishes ($V=0$), an air parcel still has non-zero absolute angular momentum (except if the storm center is on the equator).}. Given the value of the Coriolis parameter, $f$, the radial structure of angular momentum can be directly translated to the radial structure of the azimuthal wind via eq. \eqref{eq:M}. This is shown schematically in Figure \ref{fig:conceptual}a.

As air parcels spiral inwards from larger radii towards $R_{max}$, they gradually lose angular momentum due to surface friction (Figure \ref{fig:conceptual}a). Hence $M$ decreases moving radially inwards (this is also necessary in order to be inertially stable). Thus, we may understand the relationship between $R_{max}$ and $R_{17.5ms}$ in terms of the fraction of absolute angular momentum that has been \textit{lost} between $R_{17.5ms}$ and $R_{max}$: $\frac{M_{max}}{M_{17.5ms}}$. This quantity will always be less than or equal to one; it equals one only if $V_{max} = 17.5 \; ms^{-1}$ and hence the two radii are equal. Given values for $R_{17.5ms}$ and $\phi$ (with $V_{17.5ms} = 17.5 \; ms^{-1}$), we can define $M_{17.5ms}$. Given a value of $\frac{M_{max}}{M_{17.5ms}}$, we can calculate $M_{max}$. Finally, given a value of $V_{max}$ and $\phi$ we can back out $R_{max}$. The quantities $V_{max}$, $R_{17.5ms}$, and $\phi$ are routinely estimated in operations. These steps are summarized in the conceptual diagram shown in Figure \ref{fig:conceptual}b.

Mathematically, our model is defined as follows:
\begin{enumerate}
\item Calculate $M_{17.5ms}$: (Inputs: $R_{17.5ms}$ and $f$)
\begin{equation}\label{eq:Mouter}
M_{17.5ms}= R_{17.5ms}*(17.5 \; ms^{-1}) + \frac{1}{2}fR_{17.5ms}^2
\end{equation}
\item Calculate $M_{max}$: (Inputs: $M_{17.5ms}$)
\begin{equation}\label{eq:Mfrac}
    M_{max} = \left(\frac{M_{max}}{M_{17.5ms}}\right)M_{17.5ms}
\end{equation}
\item Solve for $R_{max}$ from $M_{max}$: (Inputs: $V_{max}$ and $f$).
\begin{equation}\label{eq:Rmax}
    R_{max} = \frac{V_{max}}{f}\left(\sqrt{1+\frac{2 f M_{max}}{V_{max}^2}}-1\right)
\end{equation}
\end{enumerate}
The only missing information in the model is an estimate of  $\frac{M_{max}}{M_{17.5ms}}$ in eq. \eqref{eq:Mfrac}, which we describe in the next subsection.

As an aside, we note that the above Lagrangian perspective does not imply that the radius of maximum wind itself follows a material M-surface. Indeed, radial inflow remains strong at $R_{max}$ itself, and the precise location of $R_{max}$ is not easily ascertained but rather emerges from the interplay between the radial distributions of the inward flux of angular momentum and the loss of angular momentum due to surface friction \citep{Kieu_Zhang_2017, Chen_Chavas_2020}.

\subsection{Model for $\frac{M_{max}}{M_{17.5ms}}$}

C15 provides a physical model for the complete low-level radial structure of $M$ that combines the solution of \citet{Emanuel_Rotunno_2011} for the inner deep-convecting region and the solution of \citet{Emanuel_2004} for the outer non-convecting region. The C15 model does not have a true analytic solution, as it requires two fast numerical integrations to solve for the outer solution and to match it to the inner solution. An example of the radial structure of the C15 model is shown in Figure \ref{fig:conceptual}a. As noted earlier, \citet{Chavas_Lin_2016} showed that this model can capture the characteristic modes of variability in the wind field found in nature.

% , in particular a decrease in $R_{max}$ with increasing $V_{max}$ at fixed outer size and latitude, an increase in $R_{max}$ with increasing outer size at fixed $V_{max}$ and latitude, and a slow increase in $R_{max}$ with increasing latitude at fixed $V_{max}$ and outer size.

\begin{figure*}[h]
\centerline{\includegraphics[width=0.9\textwidth]{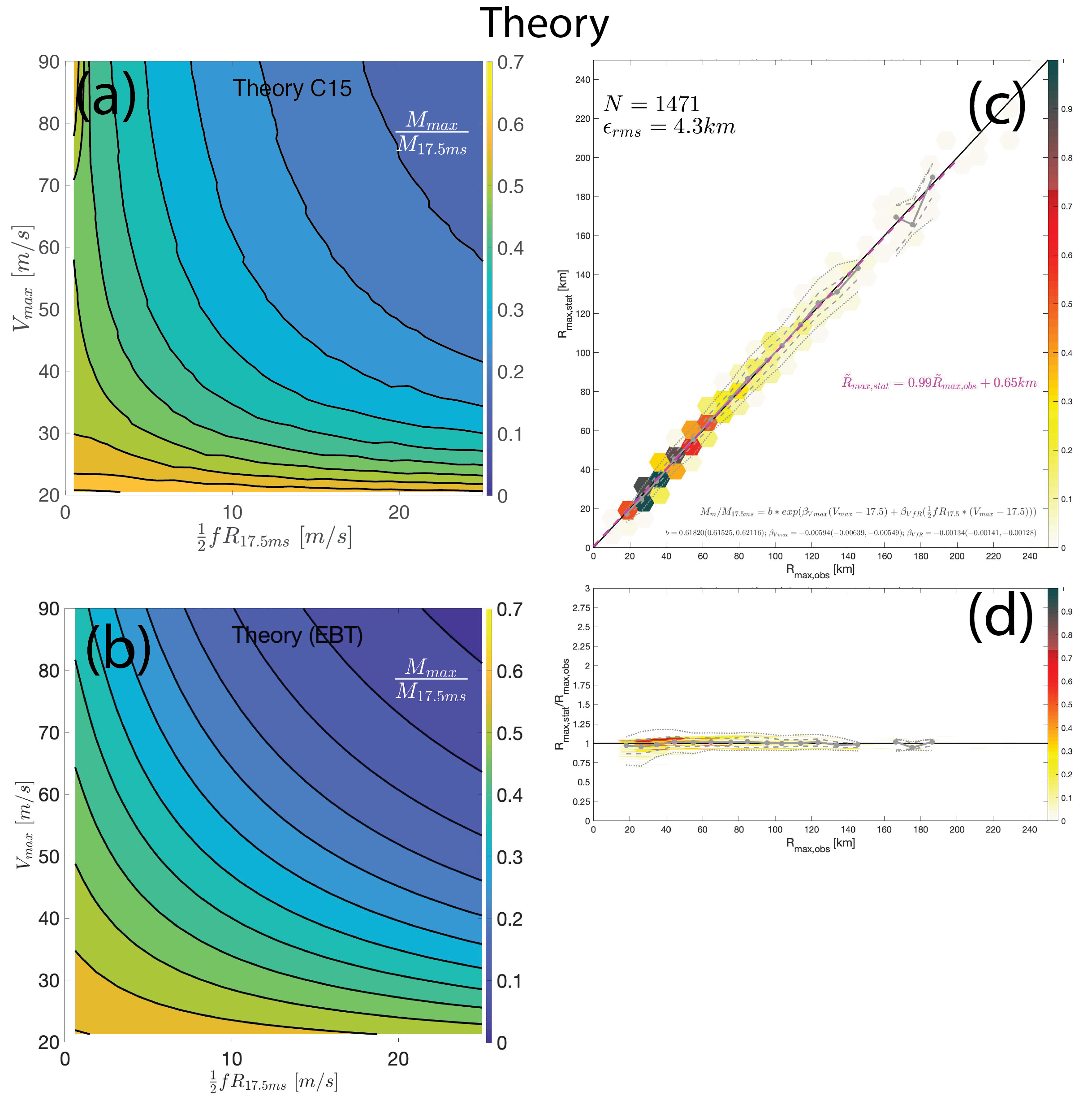}}
\caption{Theory and empirical model prediction fit to theory. (a) Theoretical distribution of angular momentum loss fraction, $\frac{M_{max}}{M_{17.5ms}}$, as a function of two velocity scales, $V_{max}$ and $\frac{1}{2}fR_{17.5ms}$, given by the hurricane wind structure model of \citet{Chavas_Lin_2016}. (b) Empirical model prediction of $\frac{M_{max}}{M_{17.5ms}}$ (eq. \eqref{eq:model_final}) fit to the theory shown in (a); fitting is performed by first interpolating the theory of (a) to the values of $\left(V_{max},\frac{1}{2}fR_{17.5ms}\right)$ in the EBT dataset for closest analog to observations. (c) Empirical model prediction of $R_{max}$ (y-axis) vs. ``observed'' $R_{max}$ (x-axis) from eqs. \eqref{eq:Mouter}-\eqref{eq:Rmax} using the model (eq. \eqref{eq:model_final}) shown in (b); color = relative frequency; black line = 1-to-1 line; gray solid/dashed/dotted lines denote conditional median, interquartile range, and 5-95\% range, respectively, within 10-km bins of the observed value starting from 0; and pink dashed line + equation = linear regression of the conditional median prediction ($\tilde{R}_{max,stat}$) vs. observed ($\tilde{R}_{max,obs}$; gray dots). (d) Same as (c) but for fractional error relative to the known value. Deviation of the slope of the pink line from the black 1-to-1 line in (c) indicates systematic bias.}
\label{fig:mfracrmax_theo}
\end{figure*}

The C15 model predicts that $\frac{M_{max}}{M_{17.5ms}}$ depends only on two physical parameters: $V_{max}$ and $\frac{1}{2}fR_{17.5ms}$. The second parameter is a velocity scale that combines information about outer size ($R_{17.5ms}$) and storm center latitude ($f$).\footnote{Theoretically, the second parameter should also be multiplied by the quantity $\frac{C_d}{w_{cool}}$, which is the ratio of the drag coefficient to the clear-sky free tropospheric subsidence rate due to radiative cooling outside of the convective inner core. This term can be neglected though as it may be taken as approximately constant from storm to storm; this assumption was also made for predicting the minimum central pressure in \citet{Chavas_Reed_Knaff_2017}.} The C15 model prediction for $\frac{M_{max}}{M_{17.5ms}}$ as a function of $V_{max}$ and $\frac{1}{2}fR_{17.5ms}$ is shown in Figure  \ref{fig:mfracrmax_theo}a. For this calculation, following C15, in the inner region we set the ratio of surface exchange coefficients $\frac{C_k}{C_d} = 1$ constant, and in the outer region we set the drag coefficient $C_d = 0.0015$ constant and the radiative-subsidence rate $w_{cool}=2 \; mm \: s^{-1}$ constant. $\frac{M_{max}}{M_{17.5ms}}$ monotonically decreases, indicating a greater loss of angular momentum, when moving from bottom-left to top-right in the figure, i.e. for higher intensities ($V_{max}$) and for higher latitude and/or outer size ($\frac{1}{2}fR_{17.5ms}$). The lone exception is at very high $V_{max}$ and very small $\frac{1}{2}fR_{17.5ms}$ (top-left corner of the plot) where the angular momentum loss fraction can actually increase with increasing intensity; we return to this briefly below.

The theory shown in Figure \ref{fig:mfracrmax_theo}a can in principle be used directly to model $\frac{M_{max}}{M_{17.5ms}}$.  However, doing so will inherently incorporate any biases in the theoretical model structure relative to that found in real storms in nature. Indeed, C15 showed that the model captures the first-order radial structure, but it also has non-negligible biases. Moreover, C15 theory lacks a simple analytic solution and must be solved numerically, which makes it less practical for everyday use.

Instead, we use the underlying theoretical basis of C15 theory in the design of a simple empirical model for $\frac{M_{max}}{M_{17.5ms}}$. We choose a log-link linear regression model for $\frac{M_{max}}{M_{17.5ms}}$ that depends on $V_{max}$ and $\frac{1}{2}fR_{17.5ms}$ as follows:
\begin{equation}\label{eq:loglinkreg}
\begin{split}
&    ln\left(\frac{M_{max}}{M_{17.5ms}}\right) = \beta_0 + \beta_{Vmax}(V_{max}-17.5 \; ms^{-1}) \\ 
& + \beta_{VfR}(V_{max} - 17.5 \; ms^{-1})\left(\frac{1}{2}fR_{17.5ms}\right) + \epsilon
\end{split}
\end{equation}
where $\beta_{Vmax}$ and $\beta_{VfR}$ are the regression coefficients to $(V_{max}-17.5 \; ms^{-1})$ and $(V_{max} - 17.5 \; ms^{-1})\left(\frac{1}{2}fR_{17.5ms}\right)$, respectively, $\beta_0$ is a constant and $\epsilon$ is the model residual error.

We can solve eq. \ref{eq:loglinkreg} for $\frac{M_{max}}{M_{17.5ms}}$ to yield our final model equation:
\begin{equation}\label{eq:model_final}
\begin{split}
&    \frac{M_{max}}{M_{17.5ms}} = \\
&    be^{\beta_{Vmax}(V_{max}-17.5 m/s) + \beta_{fR17}(V_{max} - 17.5 \; ms^{-1})\left(\frac{1}{2}fR_{17.5ms}\right)}
\end{split}
\end{equation}
where $b = e^{\beta_0}$ and we drop the error term. Model coefficients $b$, $\beta_{Vmax}$, and $\beta_{fR17}$ will be estimated from observational data. Coefficient estimation is performed using the MATLAB function 'fitglm' using the link option 'log'.

% Each coefficient can be interpreted mathematically as a fractional sensitivity of $\frac{M_{max}}{M_{17.5ms}}$ to changes in their respective predictor for a fixed value of the other predictor.

We choose a log-link model (natural logarithm on the left hand side) because $\frac{M_{max}}{M_{17.5ms}}$ is a positive definite quantity, which an exponential model always reproduces but is not guaranteed when using standard linear regression. We choose the intensity predictor to be $(V_{max}-17.5 \; ms^{-1})$ and the $\frac{1}{2}fR_{17.5ms}$ predictor to be multiplied by $(V_{max}-17.5 \; ms^{-1})$ to better capture the nonlinear dependence on the two parameters that can be seen visually in Figure \ref{fig:mfracrmax_theo}a: the sensitivity of $\frac{M_{max}}{M_{17.5ms}}$ to $\frac{1}{2}fR_{17.5ms}$ increases moving from lower to higher intensity (larger $V_{max}$); at $V_{max} = 17.5 \; ms^{-1}$, this sensitivity is zero because $\frac{M_{max}}{M_{17.5ms}} = 1$ by definition regardless of the value of $\frac{1}{2}fR_{17.5ms}$. The latter is a fundamental constraint, it is not specific to C15 theory. The form of our model explicitly builds in this non-linear dependence and further ensures that $\frac{M_{max}}{M_{17.5ms}} = 1$ at $V_{max} = 17.5 \; ms^{-1}$. These assumptions allow us incorporate the underlying physics constraining $\frac{M_{max}}{M_{17.5ms}}$ into our empirical model.

% Note that, for the intensity predictor, taking the partial derivative of eq. \eqref{eq:loglinkreg} with respect to $V_{max}$ yields
% \begin{equation}\label{eq:beta_vmax}
% \beta_{Vmax} = \frac{\partial [ln\left(\frac{M_{max}}{M_{17.5ms}}\right)]}{\partial V_{max}} = \frac{1}{\frac{M_{max}}{M_{17.5ms}}}\frac{\partial \left(\frac{M_{max}}{M_{17.5ms}}\right)}{\partial V_{max}}
% \end{equation}
% where $\beta_{Vmax}$ is the fractional sensitivity of $\frac{M_{max}}{M_{17.5ms}}$ to varying $V_{max}$ at constant $\frac{1}{2}fR_{17.5ms}$. For example, $\beta_{Vmax} = -0.01 \; (m/s)^{-1}$ translates to a 10\% decrease in $\frac{M_{max}}{M_{17.5ms}}$ for every $10 \; m/s$ increase in $V_{max}$. This interpretation is not as straightforward for $\beta_{fR17}$ unfortunately due to its non-linearity, but the sensitivity can still be calculated directly at a given value of $V_{max}$.

We first demonstrate that eq. \eqref{eq:model_final} is appropriate for modeling $\frac{M_{max}}{M_{17.5ms}}$. This is accomplished by applying eq. \eqref{eq:model_final} to C15 theory itself shown in Figure \ref{fig:mfracrmax_theo}a, and then predicting $R_{max}$. This allows us to compare the empirical predictions for $\frac{M_{max}}{M_{17.5ms}}$ and $R_{max}$ with the ``true" values given by the theory. We emphasize that the theory is \textit{not} the truth, but rather provides a reasonable first-order representation of the relationship among our structural parameters.  This offers a first test of our empirical model in which all parameters are known. To fit eq. \eqref{eq:model_final}, we calculate $\frac{M_{max}}{M_{17.5ms}}$ by interpolating the theory shown in Figure \ref{fig:mfracrmax_theo}a to the values of $\left(V_{max},\frac{1}{2}fR_{17.5ms}\right)$ in the EBT dataset discussed below; this provides the closest analog to observations. The empirical prediction of $\frac{M_{max}}{M_{17.5ms}}$ (eq. \ref{eq:model_final}) is shown in Figure \ref{fig:mfracrmax_theo}b; regression coefficients are listed in Table \ref{table:regcoeftable}.  eq. \eqref{eq:model_final} can closely reproduce the relatively complex structure found in the theory in Figure  \ref{fig:mfracrmax_theo}a. The final prediction of $R_{max}$ is compared to the theoretical value in Figure \ref{fig:mfracrmax_theo}c, and fractional errors as a function of the known $R_{max}$ are shown in Figure \ref{fig:mfracrmax_theo}d. Systematic bias is defined as the slope of the linear regression of the conditional median predicted value ($\tilde{R}_{max,stat}$) vs. observed value ($\tilde{R}_{max,obs}$); a slope of 1 indicates no systematic bias. These results demonstrate that our empirical model can reliably reproduce the ``true'' $R_{max}$ with very small error across all values of $R_{max}$, spanning a range from 10 km to over 200 km, and with nearly zero systematic bias (pink dashed line in Figure \ref{fig:mfracrmax_theo}c; linear regression slope of $0.99$). Note that the empirical model does not reproduce the increase in $\frac{M_{max}}{M_{17.5ms}}$ at very high $V_{max}$ and very small $\frac{1}{2}fR_{17.5ms}$ in the theory. Overall, this outcome indicates that the form of our empirical model for $\frac{M_{max}}{M_{17.5ms}}$ is well-suited for the task and may be applied to real data below.

We also tested two other forms of log-link regression model: 1) linear model with two predictors, $(V_{max}-17.5 m/s)$ and $\frac{1}{2}fR_{17.5ms}$; and 2) non-linear model with three predictors $(V_{max}-17.5 m/s)$, $\frac{1}{2}fR_{17.5ms}$, and $(V_{max} - 17.5 \; ms^{-1})\left(\frac{1}{2}fR_{17.5ms}\right)$. The linear model is viable, as it can capture the first-order pattern of monotonic decrease from bottom-left to top-right (Supplementary Figure S01b), but it exhibits a larger systematic bias (Supplementary Figure S01c). This larger bias arises because by definition it cannot capture the non-linear dependence on the predictors and so it misses the detailed structure in $\frac{M_{max}}{M_{17.5ms}}$ at low intensities (Supplementary Figure S01b). Thus, the added complexity of eq. \eqref{eq:model_final} compared to a linear model is both valuable for explaining variations in $\frac{M_{max}}{M_{17.5ms}}$ and useful for making predictions \citep{Shmueli_2010}. Results from the linear model will be included in the discussion below for context and comparison. In contrast, the three-predictor non-linear model is rejected because it is more complex than eq. \eqref{eq:model_final} but does not perform better in terms of error or bias. Moreover, while it can also reproduce the theoretical distribution of $\frac{M_{max}}{M_{17.5ms}}$, when it is applied to observations it produces a qualitatively different structure that is concave down rather than concave up because the coefficient of the non-linear term is found to be positive rather than negative. Thus, in the context of the results for both the linear and 2-variable non-linear model, the added complexity of the 3-variable model is at best unhelpful and may in fact be detrimental.

To summarize, our model predicts $R_{max}$ from $R_{17.5ms}$ via eqs. \eqref{eq:Mouter}-\eqref{eq:Rmax} and \eqref{eq:model_final}, whose coefficients are estimated from data in the Results section below (eq. \eqref{eq:model_final_empirical}). An optional final bias adjustment is given by eq. \eqref{eq:biasadj} below as well. The physical basis of our model offers three key advantages: 1) Choosing angular momentum loss fraction as the predictand constrains the model to be a value in the range $(0, 1]$; 2) Theory indicates that this predictand should decrease smoothly and monotonically from one towards zero; 3) This monotonic dependence can be reduced to two physical predictors. We use these advantages to define the form of our empirical model. Ultimately, the empirical basis of our model acknowledges that we do not fully understand the details of how angular momentum is lost in the inflow approaching $R_{max}$; it is undoubtedly more complex than idealized theory, and the representation and implications of these complexities remains highly uncertain. Hence, using data to directly estimate the true dependence allows us to capture the final outcome as found in nature despite the current gaps in our understanding.

\subsection{Observational data}

We test our model against data for the North Atlantic basin from the Extended Best Track v2021-03-01 \citep[][hereafter EBT]{Demuth_DeMaria_Knaff_2006}, which is by far the longest and most widely used wind structure database available. The EBT dataset provides estimates of storm location, storm intensity, storm type, wind radii (including $R_{17.5ms}$) in four quadrants (NE, NW, SE, SW), and $R_{max}$ (single value) for the period 1988-2019 in the North Atlantic. We calculate $R_{17.5ms}$ as the mean of all available non-zero values in each quadrant. For 2004 onwards, all data except $R_{max}$ are final best track data that are reanalyzed by the National Hurricane Center post-storm (Landsea and Franklin 2013); prior to 2004, wind radii data are not reanalyzed post-storm and are simply taken from the Automated Tropical Cyclone Forecast (ATCF; Sampson and Shrader 2000). All $R_{max}$ data are subjective forecaster estimates based on available aircraft and remotely-sensed data in near-real-time and are not reanalyzed post-storm. The 2021-03-01 version of EBT replaces some adivisory-based data contained in the a-deck CARQ (Combined Automated Request Query) entries with superior best track data contained in the b-deck (Galina Chirokova, personal communication 2021). Our results presented below are nonetheless quantitatively similar when using the preceding version with data only through 2018.

\begin{figure*}[h]
\centerline{\includegraphics[width=\textwidth]{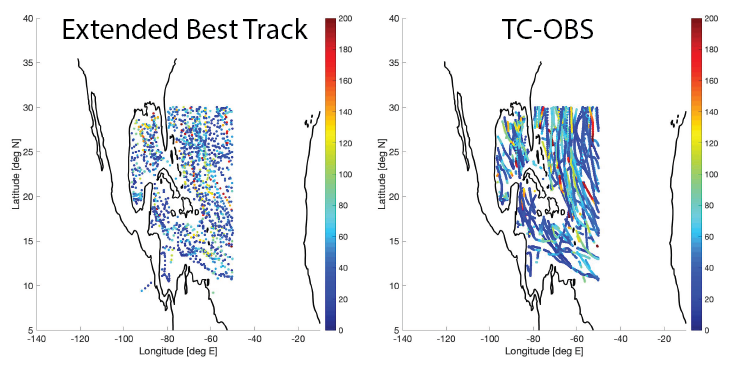}}
\caption{Maps of each dataset used in this study; color denotes $R_{max}$. Raw datasets are (a) Extended Best Track (NOAA CIRA), and (b) TC-OBS \citep[v0.40; ][]{Vigh_etal_2016}. Data filters are listed in the text.}
\label{fig:obsmapdist}
\end{figure*}

\begin{figure*}[h]
\centerline{\includegraphics[width=\textwidth]{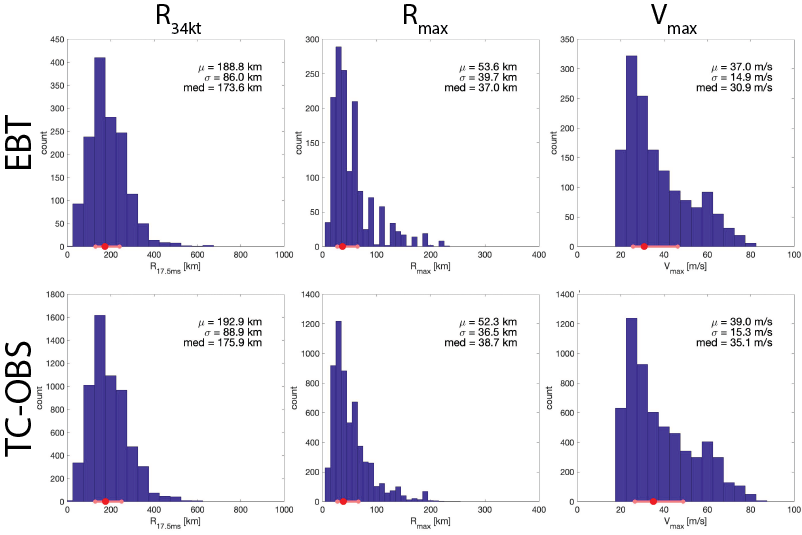}}
\caption{Histograms of $R_{17.5ms}$, $R_{max}$, and $V_{max}$ from EBT (top row) and TC-OBS (bottom row). Dots and bars along the x-axis denote the median and interquartile (25th-75th percentile) ranges.}
\label{fig:obsmapdist2}
\end{figure*}

To provide the optimal data subset for our work, we use the years 2004-2019, corresponding to the period in which the wind radii are best tracked. Moreover, we focus our analysis in the western half of the North Atlantic basin to limit ourselves to the subset of storms that were of immediate interest to forecast agencies and hence were most likely to have garnered dedicated attention from forecasters and reconnaissance \citep{Demuth_DeMaria_Knaff_2006}. Specifically, we filter the dataset as follows: 1) storm center longitude $< -50E$ (to focus on the dominant aircraft reconnaissance region); 2) $\phi < 30N$ (to minimize effects from extratropical transition); 3) storm center distance from coast $> 100 km$ (to minimize land effects on inner-core structure); 4) $0 < \frac{M_{max}}{M_{17.5ms}} \le 1$ (to remove unphysical values); 5) $6 \: km \le R_{max} \le 250 \: km$ (to remove extreme outliers); and 6) $V_{max} \ge 20 \; ms^{-1}$ (to retain storms with intensities above $17.5 \; ms^{-1}$). All data are converted to MKS units. A map of the resulting data subset is shown in Figure \ref{fig:obsmapdist}a, and histograms of $V_{max}$, $R_{max}$, and $R_{17.5ms}$ are shown in Figure \ref{fig:obsmapdist2}a-c. The final sample size is $N=1471$. This sample size is much larger than the number of degrees of freedom in our model (three: the two predictor velocities and the initial angular momentum $M_{17.5ms}$), which greatly limits the potential for overfitting.

% The operational data extends the utility of the best track information, and comes from the databases of the Automated Tropical Cyclone Forecast (ATCF; Sampson and Shrader 2000). Specifically these information come from the Combined Automated Request Query (CARQ) lines in the "aid deck" files.  Wind radii prior to 2004 are thus considered operational estimates. 

To add robustness to the choice of dataset, we perform an identical analysis with the same filters using the Tropical Cyclone Observations-Based Structure Database v0.40 \citep[][hereafter TC-OBS]{Vigh_etal_2016} for the North Atlantic. TC-OBS merges aircraft and satellite data with Extended Best Track data to provide a more objective and observationally-constrained estimate of hurricane location, intensity, and sustained near-surface wind radii and $R_{max}$ at hourly temporal resolution. The years 2004-2014 are used in our analysis to align with the start of Best Tracking of wind radii in EBT; 2014 is the final year of the latest version of the database. For storm intensity and storm central latitude, we simply use the linearly-interpolated values from the Best Track database (variables `BT\_Vmax\_interp' and `BT\_lat\_interp'). For storm structure, $R_{17.5ms}$ is calculated as the mean of all available non-zero values in each quadrant (variable `TCOBS\_wind\_radii'), while near-surface $R_{max}$ has a single value (variable `TCOBS\_maximum\_sustained\_surface\_wind\_radius'). When aircraft data are present, all wind radii and $R_{max}$ are an optimally-weighted blend of (1) the radius of the slant-adjusted flight level wind observations, (2) the radius estimated from the Stepped Frequency Microwave Radiometer (SFMR), and (3) the radius of maximum wind value from b-deck file or wind radius from the EBT database. When no nearby aircraft data are available, this parameter is relaxed toward the the interpolated b-deck value (for $R_{max}$) or EBT wind radius value with an e-folding timescale of 4-hours. A map of the resulting data subset is shown in Figure \ref{fig:obsmapdist}c, and histograms of $V_{max}$, $R_{max}$, and $R_{17.5ms}$ are shown in Figure \ref{fig:obsmapdist2}g-h. The TC-OBS subset is statistically similar to that of EBT.

Note that $R_{17.5ms}$ varies by more than a factor of 10 between smallest and largest values, whereas $f$ is restricted to vary by a factor of less than three between 10N and 30N. Hence, the parameter $\frac{1}{2}fR_{17.5ms}$ varies primarily through variations in $R_{17.5ms}$. Nonetheless, we apply our model to EBT data poleward of $30N$ below to evaluate its performance for high-latitude cases as well.

% This parameter was also found to apply to hurricanes over a much wider range of latitudes in idealized aquaplanet experiments in \citet{Chavas_Reed_Knaff_2017}.

%-----------------------
% \begin{table}[ht]
\begin{table*}[t]
\caption{Regression coefficients and statistics related to the IR-2R two regime models.  Regime 1 is for intensities less than 33 $m$ $s^{-1}$ and regime 2 for intensities above that threshold.  Predictors are current intensity ($V_{max}$), sine of storm center latitude ($sin(lat)$), IR principle component \#2 (PC2), IR-based estimate of TC size (radius of 5-kt wind, $R_{5kt}$), \% pixels colder than $-50^{o}$ $C$ ($PC50$), and $PC50^{2}$.  The intercept ($a$) and regression coefficients are provided along with the percent variance explained $R^{2}$ and the mean absolute error ($MAE$) associated with the dependent fit in km. } % title of Table
\centering % used for centering table
\begin{tabular}{c c c c c c c c c c} % centered columns (4 columns)
\hline\hline %inserts double horizontal lines
Model & a & $V_{max}$ & $sin(lat)$ & PC2 & $R_{5kt}$ & $PC50$ &$PC50^2$ & $R^2$*100 & $MAE$ \\

% inserts table
%heading
\hline % inserts single horizontal line
Regime 1 & 4.23 & -9.1$e^{-3}$ & 8.6$e^{-1}$ & 1.8$e^{-1}$ & & & & 10 & 36.7 \\ % inserting body of the table
Regime 2 & -6.67 & 9.9$e^{-3}$ & & & 7.8$e^{-2}$ & 7.7$e^{-2}$ & 1.4$e^{-4}$ & 49 & 24.8    \\ [1ex] % [1ex] adds vertical space
\hline %inserts single line
\end{tabular}
\label{table:nonlin} % is used to refer this table in the text
\end{table*}
% \end{table}
%---------------------

\subsection{Comparison and validation}

To demonstrate the utility of our model we compare the performance of our model prediction of $R_{max}$ against three existing predictive models used operationally: 1) the polynomial empirical model of \citet{Knaff_etal_2015} (their eq. 1), which depends only on $V_{max}$ and latitude; 2) the multi-satellite-platform hurricane surface wind analysis \citep[MTCSWA; ][]{Knaff_etal_2011}; and 3) a two-regime statistical method based on storm-centered infrared (IR) data (IR-2R). We test all methods against observed EBT values for the 2018-2020 hurricane seasons (the 2020 values are preliminary); for our model, we first refit the model coefficients to the EBT dataset excluding the 2018 and 2019 seasons to ensure an out-of-sample test.

The MTCSWA is a real-time surface wind analysis that combines satellite atmospheric motion vectors below 600 hPa, oceanic vector winds from scatterometry, 2-D balanced winds derived from microwave sounding instruments \citep{Bessho_etal_2004}, and an IR-based flight-level wind proxy \citep{Knaff_etal_2015} as described in \citet{Knaff_etal_2011}.  The surface winds are estimated every three hours using all of the available data, and its $R_{max}$ estimates are routinely available to operational centers.  Its routine availability and long history provide a nice comparison with a currently available technique. 

The IR-2R method has not been formally documented elsewhere, but has been running in real time since 2017. Here we briefly document how that baseline model was developed. The developmental data was divided into two regimes; 1) storms with intensities less than 33 $m$ $s^{-1}$, and 2) storms with hurricane intensities ($\geq$ 33 $m$  $s^{-1}$). The $R_{max}$ estimates used for development (1995-2014) are based on a flight-level analysis of aircraft reconnaissance flights described in \citet{Knaff_etal_2015}. The initial set of potential predictors are based on current storm characteristics ($V_{max}$ and latitude) and those derived from infrared satellite imagery that measure size, convective vigor, and radial structure. IR-based TC size is defined as an estimate of the radius of 5-kt wind ($R_{5kt}$) described in \citet{Knaff_Longmore_Molenar_2015}; convective vigor is measured by the percent pixels colder than -10, -20, -30, -40, -50, and -60 $\deg$ C.  The radial structure was estimated by azimuthally-averaged principle components of the IR brightness temperature field, also described in \citet{Knaff_Longmore_Molenar_2015}, which provide radial wavenumbers 0-4.  The regression models for both regimes are each multiple linear regression models whose predictors were determined using a leaps and bounds approach \citep{Furnival_Wilson_2000} that systematically tests all possible regressions of 1, 2, 3... variables to identify the equation with the best performance (variance explained). Variable selection is stopped when adding an additional predictor results in a reduction in explained variance less than 2\%. No independent verification or retraining was conducted, as this model is meant to be a baseline for the performance of other models. The resulting models are as follows: the regime 1 model retains a predictor related to current intensity, latitude and principle component \#2 (radial wavenumber 1).  The regime 2 model is a function of current intensity, TC size ($R_{5kt}$), the percent pixels with brightness temperatures less than $-50^{o}$ $C$ within 200 km of the TC center ($PC50$) and $PC50^{2}$. Both models are fit to the $ln$ of the $R_{max}$ and units are km. The two regimes are blended together using linear weighting between intensities of 23 and 33 $m$ $s^{-1}$. Table 1 provides the regression coefficients and statistics related to the linear fit.

% Alt: a high-resolution downscaled GFDL climate model simulation analyzed in Knutson et al. (2015)

\section{Results}\label{sec:results}

\subsection{Model results}

\begin{figure*}[h]
\centerline{\includegraphics[width=\textwidth]{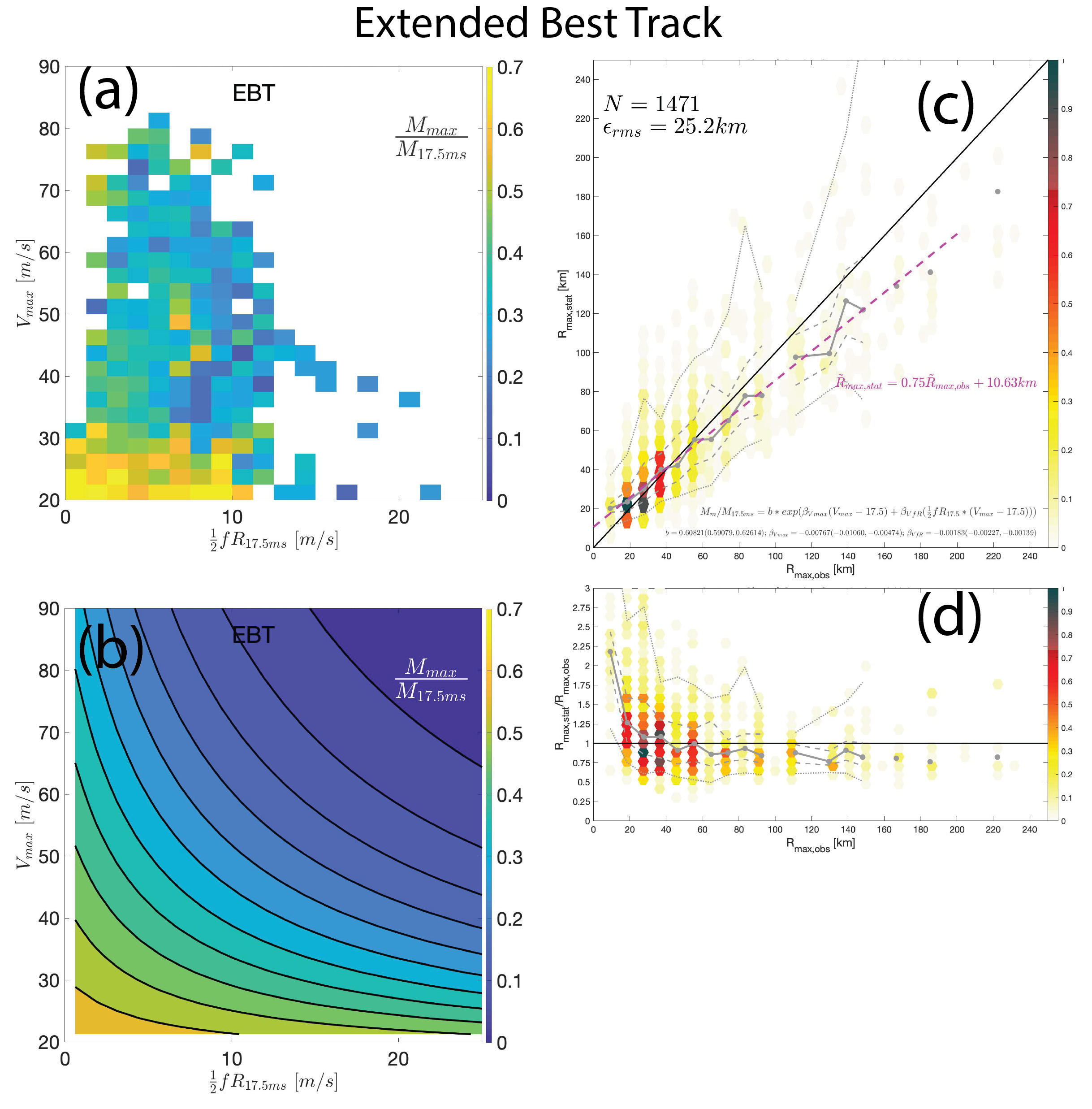}}
\caption{Empirical model prediction fit to Extended Best Track dataset for the North Atlantic 1988-2018; aesthetics as in Figure \ref{fig:mfracrmax_theo}. (a) Observed distribution of angular momentum loss fraction, $\frac{M_{max}}{M_{17.5ms}}$. (b) Empirical model prediction of $\frac{M_{max}}{M_{17.5ms}}$ fit to the data shown in (a). (c) Empirical model prediction of $R_{max}$ (y-axis) vs. observed $R_{max}$ (x-axis) using the empirical model shown in (b). (d) Same as (c) but for fractional error relative to the observed value.}
\label{fig:mfracrmax_ebt}
\end{figure*}

We now apply our model to the EBT dataset. The observed distribution of $\frac{M_{max}}{M_{17.5ms}}$ (eq. \eqref{eq:Mouter}) is shown in Figure \ref{fig:mfracrmax_ebt}a. $\frac{M_{max}}{M_{17.5ms}}$ tends to decrease towards higher values of $V_{max}$ and/or $\frac{1}{2}fR_{17.5ms}$, consistent with theory. The regression model fit to the observed data is shown in Figure \ref{fig:mfracrmax_ebt}b, which is given by the equation
\begin{equation}\label{eq:model_final_empirical}
\begin{split}
&    \frac{M_{max}}{M_{17.5ms}} = \\
&   0.608e^{-0.00767(V_{max}-17.5 m/s) - 0.00183(V_{max} - 17.5 \; ms^{-1})\left(\frac{1}{2}fR_{17.5ms}\right)}
\end{split}
\end{equation}
The regression model coefficients are provided in Table \ref{table:regcoeftable}. The coefficients for the two predictors are negative, which provides quantitative confirmation that $\frac{M_{max}}{M_{17.5ms}}$ decreases statistically for higher values of each predictor.  The dependence on $\frac{1}{2}fR_{17.5ms}$ has wider uncertainties, which arises in part because $\frac{1}{2}fR_{17.5ms}$ has a smaller range of values than $V_{max}$, but also indicates more variability in the data. The structure of the dependence of the model fit to EBT shown in Figure \ref{fig:mfracrmax_ebt}b looks remarkably similar to that for the theory shown in Figure \ref{fig:mfracrmax_theo}b. Indeed, the EBT coefficients are qualitatively similar to theory, but they are quantitatively a bit different: this is precisely the bias in the theory that we avoid by using an empirical model with coefficients estimated directly from data.

%---------------------
% \begin{table}[ht]
\begin{table*}[t]
\caption{Regression coefficients for our log-link nonlinear regression model (eq. \ref{eq:model_final}) fit to theory, Extended Best Track, and TC-OBS. 95\% confidence intervals are shown in parentheses. See text for details.} % title of Table
\centering % used for centering table
\begin{tabular}{c c c c c} % centered columns (4 columns)
\hline\hline %inserts double horizontal lines
coefficient & Theory & Extended Best Track & TC-OBS \\ [0.5ex] % inserts table
%heading
\hline % inserts single horizontal line

$\beta_{Vmax}$ [m/s]    & -0.00594 (-0.00639,-0.00549)    &  -0.00767 (-0.01060,-0.00474)   & -0.00727 (-0.00860,-0.00594)    \\ % inserting body of the table
$\beta_{VfR}$ [m/s]     & -0.00134 (-0.00141,-0.00128)    &  -0.00183 (-0.00227,-0.00139)   & -0.00161 (-0.00180,-0.00143) \\
$b$ [m/s]               & 0.618 (0.615,0.621)    &  0.608 (0.591,0.626)      & 0.630 (0.620,0.640) \\
% Diagnostic &   -2.4 & 34.8 & 37 \\ [1ex] % [1ex] adds vertical space
\hline %inserts single line
\end{tabular}
\label{table:regcoeftable} % is used to refer this table in the text
\end{table*}
% \end{table}
%---------------------

The final prediction of $R_{max}$ is compared to the theoretical value in Figure \ref{fig:mfracrmax_ebt}c, and fractional errors as a function of the observed $R_{max}$ are shown in Figure \ref{fig:mfracrmax_ebt}d. Model error and bias are presented in Table \ref{table:modelcompare}. The model can consistently capture the first order variability in $R_{max}$ over a wide range of values from 15-200 km. There is a slight systematic bias in the prediction of $R_{max}$ (linear regression slope of $0.75$) in which the model overestimates $R_{max}$ at small values and underestimates it at large values, with the transition point at approximately 40 km (pink dashed line). Above this threshold, the systematic underestimation is roughly constant at approximately 20\% of the observed $R_{max}$. Below this threshold, overestimation can increase to very large values as $R_{max}$ becomes very small; this is unsurprising given that estimates in the observed $R_{max}$ itself carry substantial uncertainty whose relative errors will be magnified as $R_{max}$ becomes very small. This uncertainty is at least in part due to the tendency to discretize operational estimates into 5 nautical mile bins, though it may also indicate real physical variability within compact inner cores of hurricanes. Moreover, there is evidence of slight overestimation of $R_{max}$ in the EBT \citep{Combot_etal_2020}.

Table \ref{table:modelcompare} also compares our model performance statistics against both the linear version of our model, with predictors $(V_{max} - 17.5 \; ms^{-1})$ and $\frac{1}{2}fR_{17.5ms}$, as well as the \citet{Knaff_etal_2015} model. The nonlinear model yields a lower systematic bias than the linear model (Supplementary Figure S02c). As noted earlier, the linear model can capture the qualitative pattern of $\frac{M_{max}}{M_{17.5ms}}$ but the linear dependence on the predictors prevents it from capturing the non-linear structure at lower intensities (Supplementary Figure S02b). Our model also performs substantially better than the model of \citet{Knaff_etal_2015}, whose conditional best-fit slope is only marginally higher than zero, indicating strong systematic bias, due in part to the fact that it predicts a relatively limited range of variability in $R_{max}$ between 25 and 70 km.

%---------------------
% \begin{table}[ht]
\begin{table*}[t]
\caption{Comparison of model performance statistics. For systematic bias, a slope of one is unbiased.} % title of Table
\centering % used for centering table
\begin{tabular}{c c c c} % centered columns (4 columns)
\hline\hline %inserts double horizontal lines
Model& RMSE [km] & Systematic bias (slope) \\ [0.5ex] % inserts table
%heading
\hline % inserts single horizontal line
Our model (eq. \eqref{eq:model_final_empirical})     &  25.2     & 0.75  \\ % inserting body of the table
Two-predictor linear log-link         & 21.9      & 0.66  \\
\citet{Knaff_etal_2015} eq. 1                & 39.9    & 0.11 \\ [1ex] % [1ex] adds vertical space
\hline %inserts single line
\end{tabular}
\label{table:modelcompare} % is used to refer this table in the text
\end{table*}
% \end{table}
%---------------------

% [plot of error statistics (Rmaxpred/Rmax) vs. Vmax? vs R17.5ms?]

\begin{figure*}[h]
\centerline{\includegraphics[width=\textwidth]{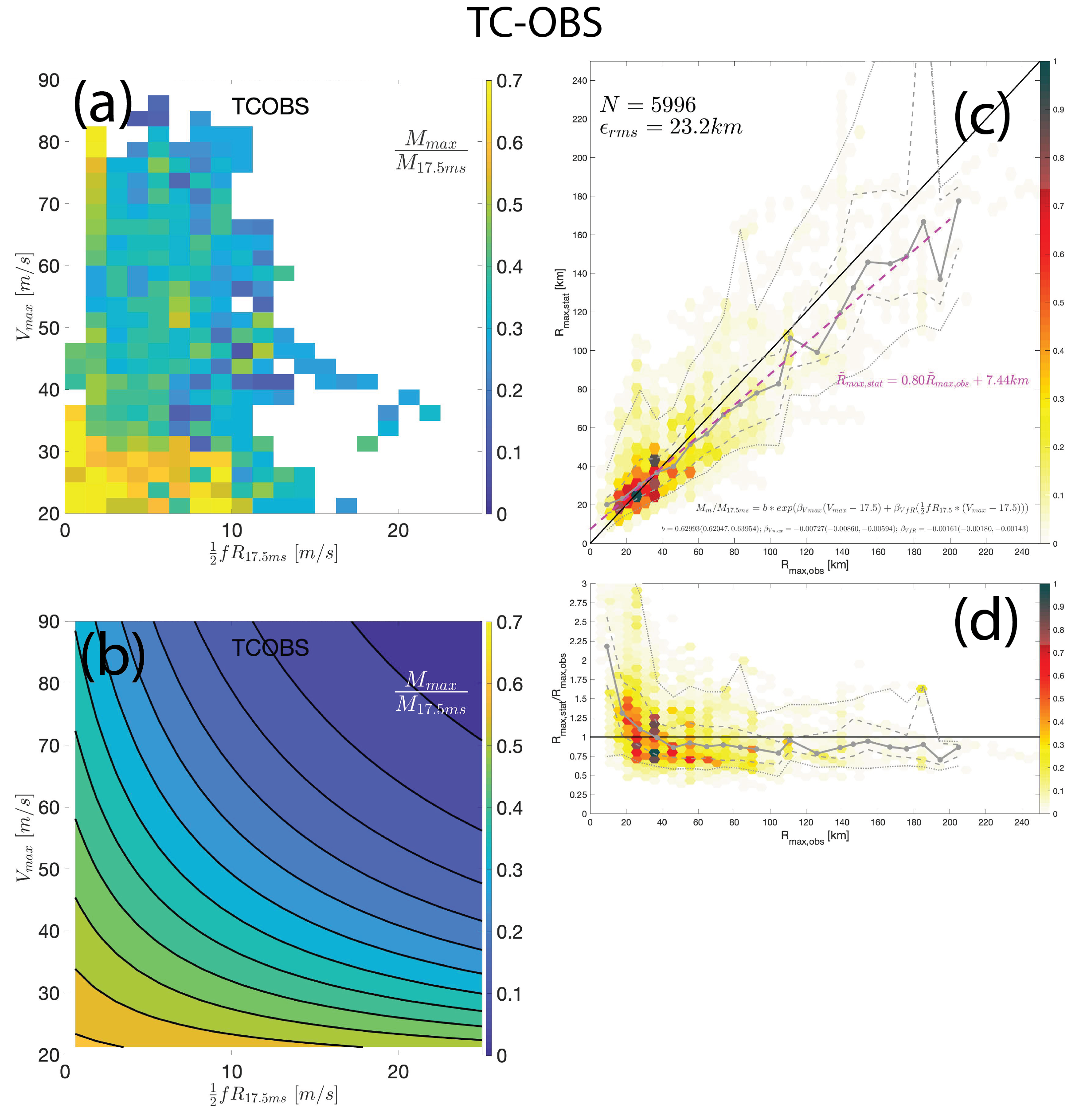}}
\caption{Empirical model prediction fit to TC-OBS dataset for the North Atlantic 2004-2014; aesthetics as in Figure \ref{fig:mfracrmax_ebt}. Results are quantitatively similar to the those for EBT shown in Figure \ref{fig:mfracrmax_ebt}.}
\label{fig:mfracrmax_tcobs}
\end{figure*}

Quantitatively similar results to EBT are found in TC-OBS as shown in Figure \ref{fig:mfracrmax_tcobs}. The empirical model structure and coefficient estimates are quantitatively similar to those of EBT (Table \ref{table:regcoeftable}), and the final prediction of $R_{max}$ yields similar errors and systematic biases. This result lends greater confidence in the robustness of our empirical modeling results.

% %---------------------
% \begin{table}[ht]
% \caption{Regression coefficients for log-link linear regression model with predictors $V_{max} - 17.5 \; ms^{-1}$ and $\frac{1}{2}fR_{17.5ms}$, for comparison to our non-linear model in Table \ref{table:regcoeftable}.} % title of Table
% \centering % used for centering table
% \begin{tabular}{c c c c c} % centered columns (4 columns)
% \hline\hline %inserts double horizontal lines
% coefficient & Theory & Extended Best Track & TC-OBS \\ [0.5ex] % inserts table
% %heading
% \hline % inserts single horizontal line

% $\beta_{Vmax}$ [m/s]    & -0.0059 ()    &  -0.0077 ()   & -0.0073 ()    \\ % inserting body of the table
% $\beta_{fR}$ [m/s]     & -0.0013 ()    &  -0.0018 ()   & -0.0016 () \\
% $b$ [m/s]               & 0.618 ()    &  0.608 ()      & 0.630 () \\
% % Diagnostic &   -2.4 & 34.8 & 37 \\ [1ex] % [1ex] adds vertical space
% \hline %inserts single line
% \end{tabular}
% \label{table:regcoeftable} % is used to refer this table in the text
% \end{table}
% %---------------------

%Quantitative parameter estimates are quite similar for $V_{max}$ but vary more substantially for $\frac{1}{2}fR_{17.5ms}$, though those differences are not necessarily statistically significant due to the inherent uncertainty in the estimate of that parameter as noted above.

One notable deviation from the empirical model is the apparent slight \textit{increase} in $\frac{M_{max}}{M_{17.5ms}}$ moving towards very high intensities at small values of $\frac{1}{2}fR_{17.5ms}$ evident in both the EBT (Figure \ref{fig:mfracrmax_ebt}a) and TC-OBS (Figure \ref{fig:mfracrmax_tcobs}a) databases. This behavior is not predicted by the empirical model presented here, but it does show up in a similar portion of the phase space in the C15 theory in Figure \ref{fig:mfracrmax_theo}a as noted above. Physically, this behavior indicates that at very high intensity and for small and/or high-latitude storms, less angular momentum is lost to the surface between $R_{17.5ms}$ and $R_{max}$ at higher intensity than at lower intensity. More generally, this result may suggest an important change in the qualitative behavior of the physics of the boundary layer and its interaction with the surface, which may be due to e.g. a reduction in the drag coefficient at high wind speeds, which is a topic of ongoing debate \citep{Richter_etal_2021, Richter_Stern_2014, Donelan_etal_2004}. Here our analysis suggests such effects occur specifically at very small $\frac{1}{2}fR_{17.5ms}$. Deeper interpretation of this result and its manifestation in C15 theory lies well beyond the scope of this work. However, we emphasize here that the quantity $\frac{M_{max}}{M_{17.5ms}}$ represents the radially-integrated effect of surface drag on an air parcel, which may hold useful information for helping to constrain the variation of $C_d$ in the inner core of a hurricane.

% [stratify intensify/weaken + table]

% In our case, observed values of $R_{max}$ are skewed strongly towards smaller values where there is higher intrinsic uncertainty as noted above, and thus the empirical model will be weighted towards reproducing those values at the expense of the higher values. 

\subsection{Application to a historical case}

\begin{figure*}[h]
\centerline{\includegraphics[width=\textwidth]{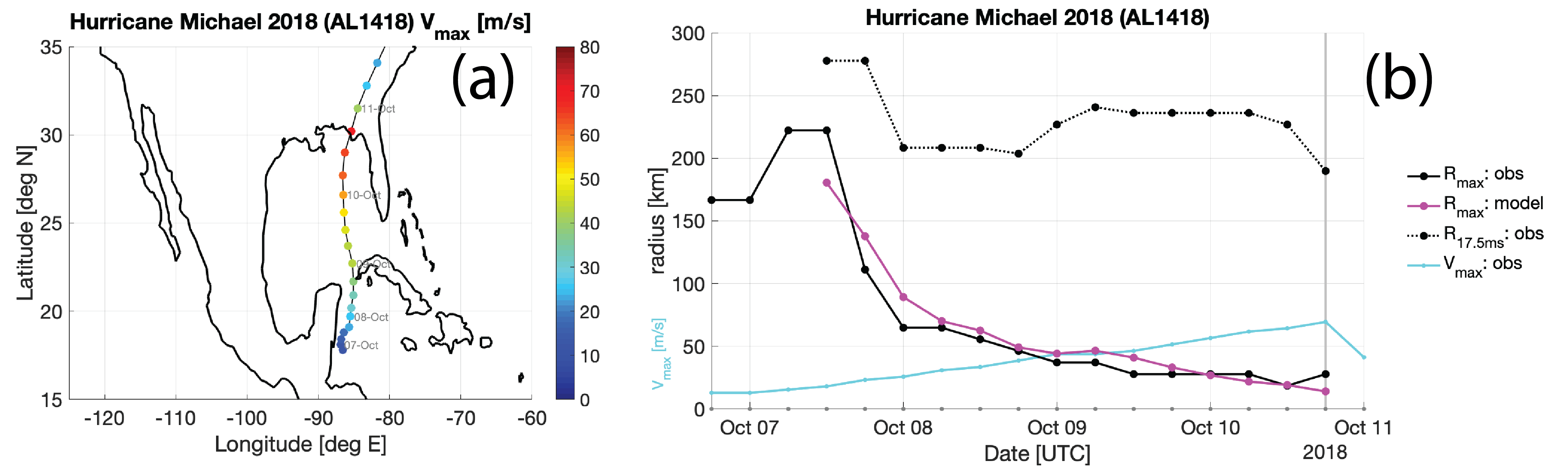}}
\caption{Example application of our model to Hurricane Michael (2018). (a) Map of storm track with Best Track $V_{max}$ (color dots). (b) Predicted $R_{max}$ (pink), EBT $R_{max}$ (black solid), EBT $R_{17.5ms}$ (black dashed) and $V_{max}$ (light blue) prior to landfall (light grey line).}
\label{fig:model_example}
\end{figure*}

An example application of the model to a historical TC is presented in Figure \ref{fig:model_example} for the life-cycle of Hurricane Michael (2018), which formed at 1800 UTC on October 6 2018 and made landfall in Florida as a Category 5 storm at 1800 UTC on October 10. A map of Michael's track and intensity is shown in Figure \ref{fig:model_example}a. Michael gradually and monotonically intensified during its lifecycle leading up to landfall. We compare the observed $R_{max}$ against the prediction from our model in Figure \ref{fig:model_example}b. Michael's $R_{max}$ was initially greater 200 km and then decreased rapidly to 64 km over the 12-hour period from 1200 UTC on October 7 to 0000 UTC on October 8 before decreasing gradually with time until landfall; this evolution is predicted well by our model. During the early 12-hour period of rapid decrease in $R_{max}$, storm outer size ($R_{17.5ms}$) also decreased significantly from 278 km to 208 km while storm intensity ($V_{max}$) increased from 18 m/s to 26 m/s. Thus, $R_{max}$ is predicted to decrease due to both of these effects occurring simultaneously: contraction of the inner core with intensification and shrinking of the storm as a whole. Thereafter, $R_{max}$ decreases gradually with time as the storm gradually intensifies, whereas outer size ($R_{17.5ms}$) remains relatively constant with values between 208 km and 241 km and $f$ increases only modestly as Michael moves northward. Thus, $R_{max}$ is predicted to decrease principally due to the increase in intensity. Note that $R_{max}$ changes much more rapidly with changes in intensity at lower intensity, a behavior captured by our model.

\subsection{Application to higher latitude storms}

\begin{figure*}[h]
\centerline{\includegraphics[width=0.5\textwidth]{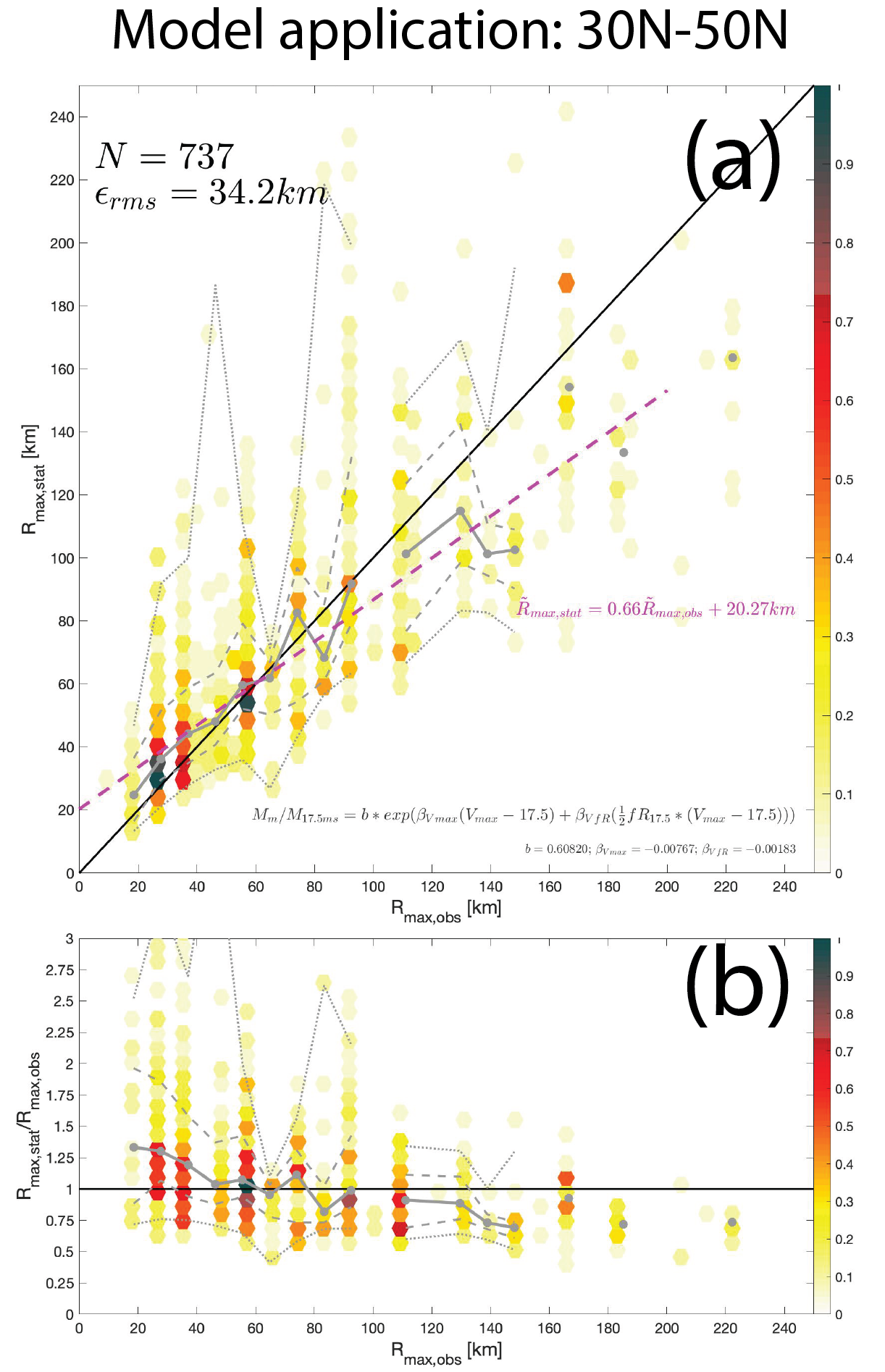}}
\caption{As in Figure \ref{fig:mfracrmax_ebt}c-d but for our model prediction of $R_{max}$ (from eq. \eqref{eq:model_final_empirical}) applied to North Atlantic EBT data between 30N and 50N.}
\label{fig:highlat}
\end{figure*}

It is useful to further evaluate how our model performs when applying it to data poleward of 30N. Though we trained the model equatorward of 30N to avoid the messier details associated with extratropical interactions, the theory is in principle valid at any latitude. Figure \ref{fig:highlat} shows the result of applying eq. \eqref{eq:model_final_empirical} to our EBT dataset filtered in the same way as above except now for the latitude band 30-50N. The model performs reasonably well overall given the greater uncertainties both observationally and theoretically at these higher latitudes, with slightly greater RMS error (34.2 km) and systematic bias (conditional slope of 0.66). The transition from overestimate to underestimate occurs at an observed value of approximately 60 km, with median fractional errors towards higher observed values again remaining relatively constant at an underestimate of approximately 25\%. These results suggest that the model is suitable for application at higher latitudes as well.

\subsection{$R_{max}$ vs. latitude}

\begin{figure*}[h]
\centerline{\includegraphics[width=\textwidth]{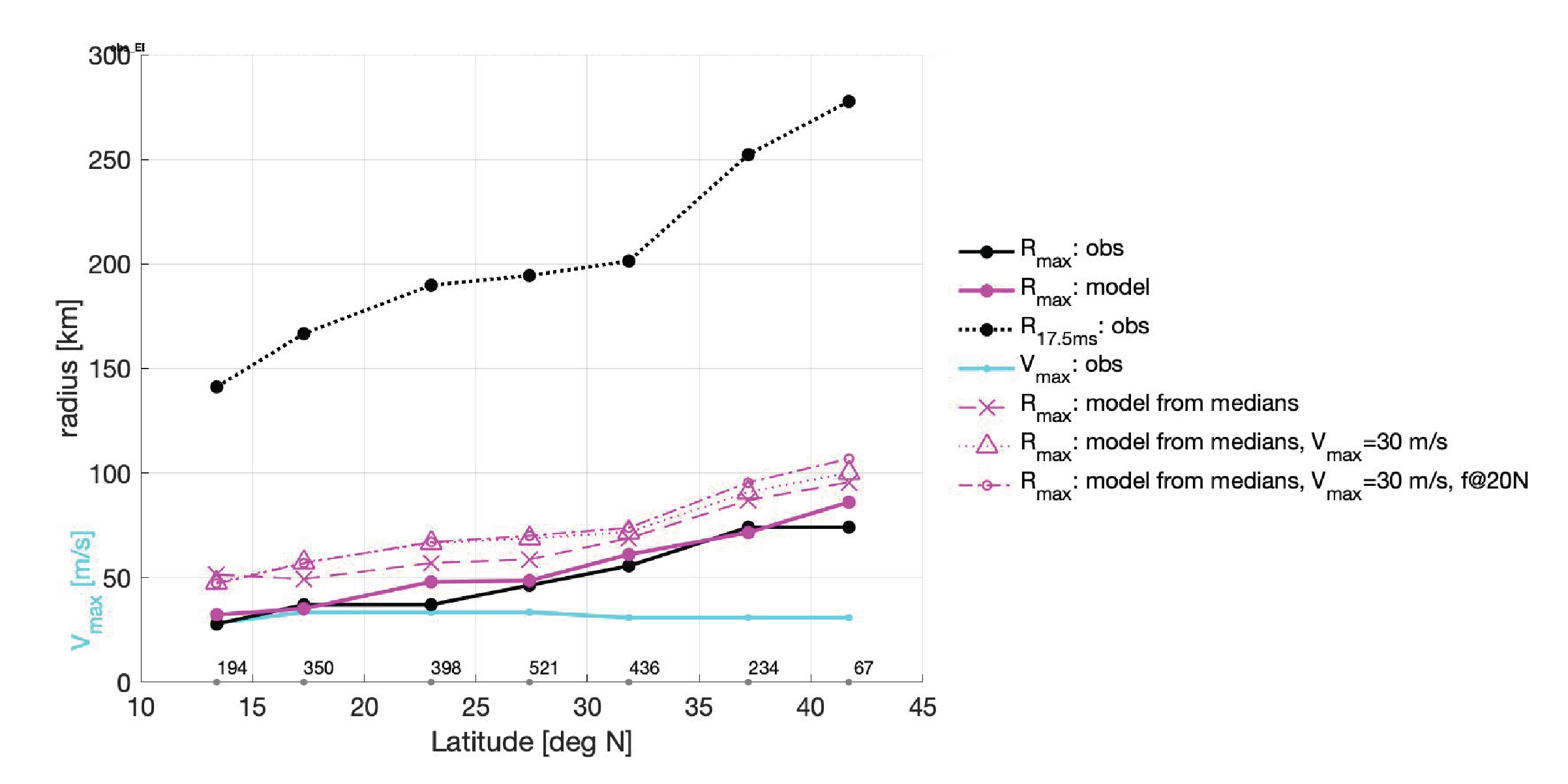}}
\caption{The model predicts the statistical increase in $R_{max}$ with latitude found in observations. Median values of predicted $R_{max}$ (pink), EBT $R_{max}$ (black solid), EBT $R_{17.5ms}$ (black dashed) and $V_{max}$ (light blue) vs. median latitude in five-degree latitude bins; lowest latitude bin includes all data south of 15N and highest latitude bin includes all data north of 40N. Also shown: model prediction direct from bin-median values of latitude, $V_{max}$, and $R_{17.5ms}$ (pink x); this same prediction but holding $V_{max} = 30 \; ms^{-1}$ constant (pink triangle); further holding $f$ constant at its value at 20N (pink circle). The latter tests demonstrate how the statistical increase in $R_{max}$ with latitude is due to the increase in $R_{17.5ms}$ with latitude.}
\label{fig:rmax_vs_latitude}
\end{figure*}

Statistical changes in $R_{max}$ and other storm parameters with latitude are shown in Figure \ref{fig:rmax_vs_latitude}. Observed $R_{max}$ increases with latitude, as has been noted in past studies, from 28 km south of 15N to 74 km north of 35N (solid black line). This behavior is quantitatively well-captured by our model (solid pink line), including poleward of 30N.

For the input parameters, median $R_{17.5ms}$ increases substantially with latitude (141 km south of 15 N; 278 km north of 40N), whereas median $V_{max}$ is nearly constant with latitude with a slight decrease from $33.4 \; ms^{-1}$ within 15-30N to $30.9 \; ms^{-1}$ north of 30N. This suggests that $R_{max}$ tends to increase statistically with latitude principally because $R_{17.5ms}$ increases significantly with latitude. We test this hypothesis explicitly using our model. First, our model can reproduce the increase in latitude in a simpler manner by applying it directly to the median values of each input parameter within each bin (pink x's). This yields a similar increase with latitude; the values are systematically biased high relative to the true median within each bin, an indication of the non-linearity of the problem. We then perform the same model prediction from median values but holding $V_{max}=30 \; ms^{-1}$ constant (pink triangles). The result is a very similar increase in $R_{max}$ with latitude as before, indicating that the small decrease in $V_{max}$ with latitude is not important for this behavior. Finally, we further hold $f$ constant at its value at 20N (pink circles) to fully isolate the effect from the increase in outer size alone. This result is only a modest effect on the trend with latitude, indicating that variations in $f$ are also not important. Taken together, then, $R_{max}$ increases with latitude predominantly because outer size ($R_{17.5ms}$) increases with latitude.

% Finally, this behavior is not intrinsic to the model itself. If we take $V_{max} = 30 \; ms^{-1}$ constant and $R_{17.5ms} = 200 \; km$ constant, our model predicts $R_{max}$ will stay relatively constant: $R_{max} = 68 \; km$ at 10N, $R_{max} = 71 \; km$ at 25N, and $R_{max} = 72 \; km$ at 40N. This very slight increase with latitude reverses to a slight decrease with latitude at a higher intensity of $V_{max} = 50 \; ms^{-1}$.

%to output from simulated storms analyzed in Knutson et al. (2015) and Chavas et al (2017) to test a physical-statistical prediction for the radius of maximum wind from outer storm size.

\subsection{Comparison with existing predictive models}

We next compare out-of-sample predictions from our model against three existing operational models: the \citet{Knaff_etal_2015} model, MTCSWA, and IR-2R. We use the EBT datasets for test period of 2018-2020 using the same filters as were applied above. To provide a true out-of-sample test of our model, we refit our model coefficients to the EBT dataset excluding the years 2018-2019. The resulting coefficients ($\beta_{Vmax} = -0.00701$, $\beta_{VfR}=-0.00181$, $b=0.600$) are very similar to the values for the full dataset (Table \ref{table:regcoeftable}). Because this sample size is much smaller, we define systematic bias using a simple linear regression fit to the predicted vs. observed $R_{max}$.

\begin{figure*}[h]
\centerline{\includegraphics[width=0.9\textwidth]{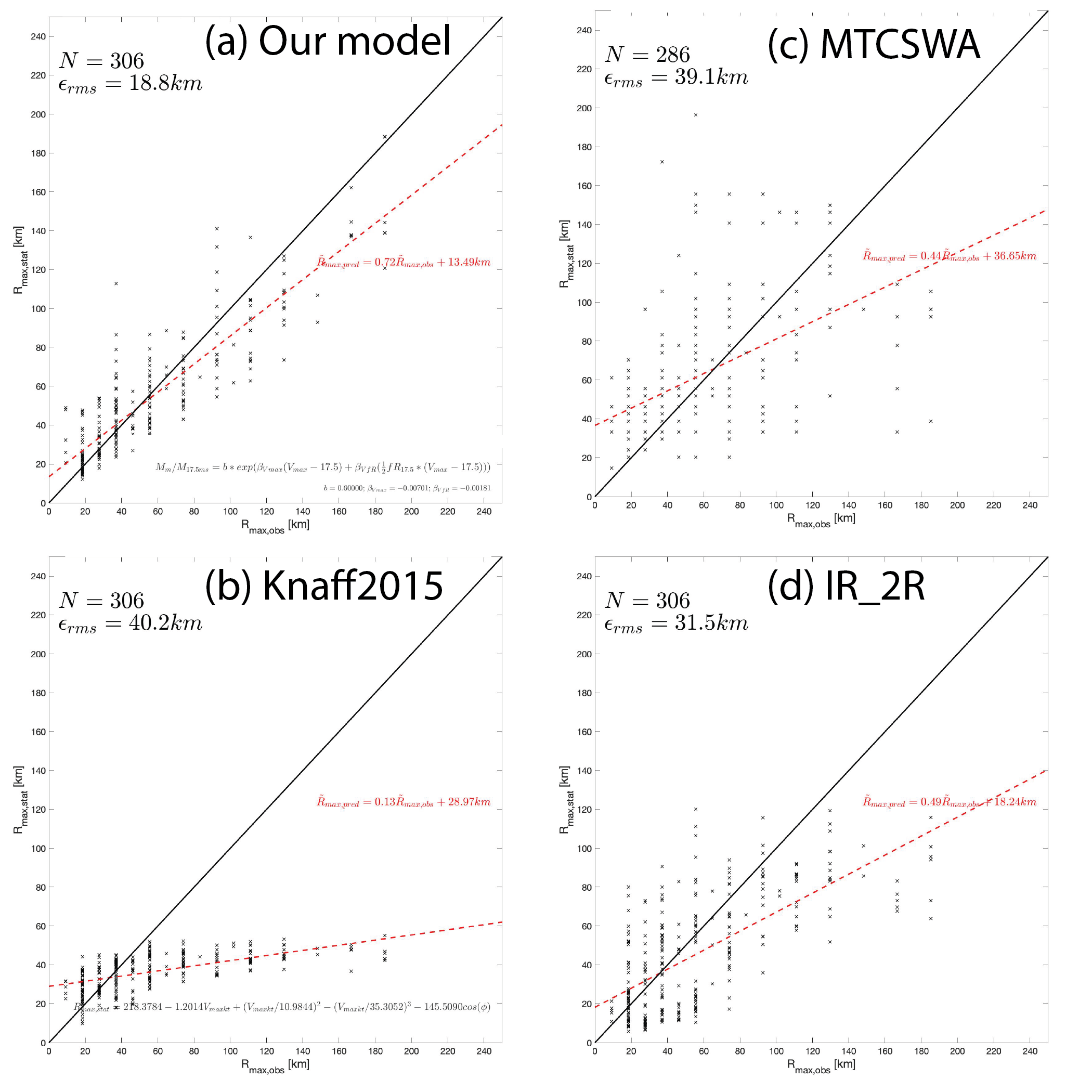}}
\caption{Scatter plots of out-of-sample predicted vs. observed $R_{max}$ for EBT data 2018-2020. (a) Our model (coefficients refit to EBT excluding 2018-2019); (b) \citet{Knaff_etal_2015} model; (c) MTCSWA; and (d) IR-2R. Black line = 1-to-1 line, and red dashed line = linear fit to data.}
\label{fig:compareremote}
\end{figure*}

Overall, our model performs substantially better than all three existing predictive models (Figure \ref{fig:compareremote}). Our model has an out-of-sample RMS error of 18.8 km and systematic bias slope of 0.72, similar to the full-dataset results of Figure \ref{fig:mfracrmax_ebt}. As found above for the full dataset (Table \ref{table:modelcompare}), the \citet{Knaff_etal_2015} model has relatively little predictive power. Meanwhile,  MRCSWA and IR-2R both exhibit substantially larger RMS errors (39.1 km and 31.5 km, respectively) and systematic biases (0.44 and 0.49, respectively) than our model.

% are made with $R_{max}$ estimated on inbound and outbound radial legs of NOAA and U.S. Air Force reconnaissance aircraft as described in section 2.  The match-ups are considered if the aircraft observation is within 2 h of the real-time estimate and only the last several years (2018-2020) are considered in order to ensure these are independent of the methods developed. 

% Summary statistics including mean errors, root mean square errors, and the percent variance explained for each model are contained in Table 2. It is recognized that $R_{max}$ is often a difficult quantity to estimate due to the highly variable nature of the surface winds and incomplete sampling. But, our problem is more difficult; estimating the poorly-observed $R_{max}$ when surface wind observations are unavailable.  The summary statistics reflect both the quality of the observational estimates and of the difficult nature of the problem.  In this short, but independent dataset, $RMSEs$ range from 44 to 34 $km$, and biases range from -10 to 10 $km$.  The percent variances also give us an idea of the goodness of fit with the IR-2R model  explaining the least, 26\% and the Diagnostic model developed here 37 \%.  

% We now examine the scatter plots of these estimates in Figure 8.

Between the two remote sensing-based models, there is an indication that the training data likely did not have as large of a range of possibilities that were observed in the past three years.  IR-2R does not predict $R_{max}$ values much larger than about 120 $km$, though it does have a smaller error than MTCSWA.  The MTCSWA analysis, on the other hand, has much larger predicted values though often associated with moderate rather than large values of $R_{max}$ and hence yield large errors.

\subsection{Optional final bias adjustment}

\begin{figure*}[h]
\centerline{\includegraphics[width=0.5\textwidth]{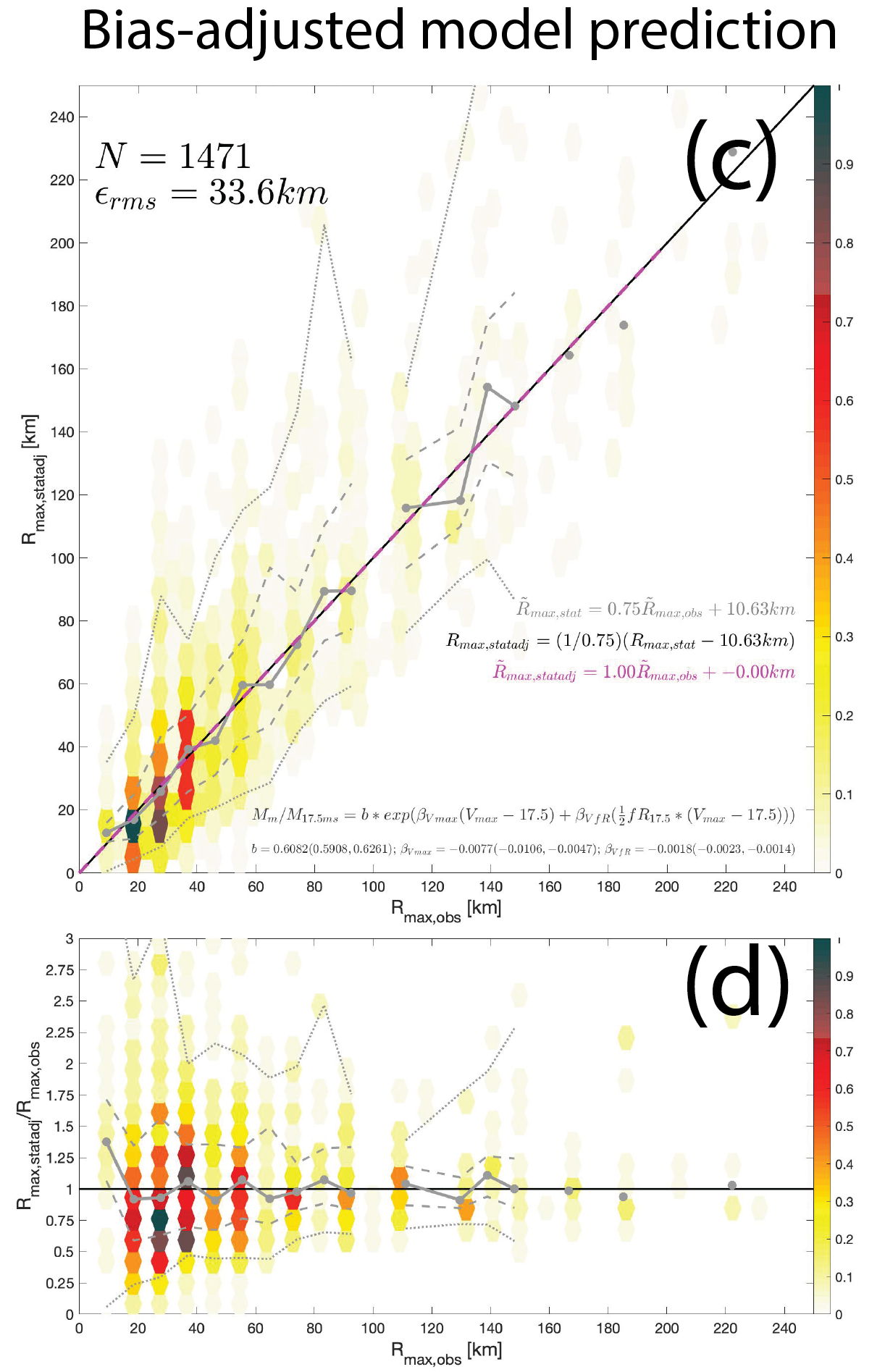}}
\caption{As in Figure \ref{fig:mfracrmax_ebt}c-d but adjusted to remove systematic bias in prediction of $R_{max}$. Bias adjustment is given by eq. \eqref{eq:biasadj}.}
\label{fig:mfracrmax_ebt_biasadj}
\end{figure*}

For the purposes of prediction, we may take one final step and adjust our model prediction to remove the systematic bias. This is done by solving the equation for the linear regression on the conditional median shown in pink in Figure \ref{fig:mfracrmax_ebt}c for $\tilde{R}_{max,obs}$, which we will denote here as our bias-adjusted final prediction $R_{max,statadj}$:
\begin{equation}\label{eq:biasadj}
    R_{max,statadj} = \frac{1}{0.75}(R_{max,stat} - 10.63 \: km)
\end{equation}
Doing so eliminates the systematic bias (Figure \ref{fig:mfracrmax_ebt_biasadj}), though it slightly increases the overall RMSE error to 33.1 km. This outcome is expected: by definition, this multiplicative adjustment will increase the range of predicted values within each observational bin, and in particular will increase absolute errors in overestimated predictions more strongly than it will decrease the absolute errors in underestimated predictions.  Such behavior is an example of the ``bias-variance trade-off'', in which adding this additional complexity to our model reduces bias but amplifies errors associated with noise in a dataset \citep{Shmueli_2010}. Since the goal of our model is to provide a reasonable first-order prediction of $R_{max}$ across a wide range of values, this bias adjustment may be desirable even at the expense of a modest increase in RMS error.

Ultimately, this bias adjustment has no obvious physical explanation and is solely a means of improving the final prediction. It may be most useful for risk prediction models where the objective is to produce $R_{max}$ values that are statistically similar to observations rather than make predictions of individual values directly from noisy observations.

\section{Conclusions}\label{sec:conclusions}

The hurricane radius of maximum wind ($R_{max}$) is critical for estimating the magnitude and footprint of wind, surge, and rainfall hazards. However, $R_{max}$ is noisy in observations and poorly resolved in reanalyses and climate models. In contrast, wind radii from the outer circulation are much less sensitive to such issues. The radius of 34-kt wind ($R_{17.5ms}$) is the outermost wind radius that is routinely estimated operationally and has been Best Tracked since 2004.

Here we have presented a simple model for predicting $R_{max}$ from $R_{17.5ms}$ that combines the underlying physics of hurricane radial structure with empirical data. The model is given by eqs. \eqref{eq:Mouter}-\eqref{eq:Rmax} and \eqref{eq:model_final_empirical}, with an optional final bias adjustment given by eq. \eqref{eq:biasadj}. Our approach uses the physics of angular momentum loss in the hurricane inflow to reduce the relationship between the two radii to a dependence on two simple physical parameters and then uses observational data to estimate the model coefficients. The form of the empirical model is chosen purposefully to incorporate the mathematical constraints that are imposed by the physics of our problem.

Our findings are summarized as follows:
\begin{enumerate}
    \item Our model offers a promising first-order prediction of $R_{max}$ from outer size (here $R_{17.5ms}$) in real hurricanes. The model only requires as input quantities that are routinely estimated operationally ($V_{max}$, $R_{17.5ms}$, and storm center latitude).
    \item Model results are consistent between Extended Best Track and TC-OBS historical hurricane databases for the North Atlantic.
    \item Model performance exceeds existing operational models for predicting $R_{max}$.
    \item The model predicts the observed statistical increase in $R_{max}$ with latitude and demonstrates that this is principally driven by the statistical increase in $R_{17.5ms}$ with latitude.
\end{enumerate}

Our model is fast and straightforward to implement, and it has value for a range of applications. The model can provide simple, observationally-constrained first-order estimates of $R_{max}$ for hurricane forecasting. This estimate could then be further refined in the presence of additional data as needed. The simplicity of the model may further allow one to quantify how uncertainties in observational estimates of input parameters, such as $R_{34kt}$, translate to uncertainties in the prediction of $R_{max}$. The model can also be used to estimate $R_{max}$ for hurricane risk applications using hurricane track models that are either empirical \citep{Yonekura_Hall_2011} or downscaled from climate models \citep{Emanuel_Ravela_Vivant_Risi_2006,Jing_Lin_2020,Lee_etal_2018}. Given $V_{max}$ and $R_{max}$, a simple parametric model (e.g. the C15 model presented here or that of \citet{Willoughby_Rahn_2004}) may be used to model the entire hurricane wind field in a manner that optimally captures the strongest winds in the vicinity of $R_{max}$, which is ideal for hazard modeling. Finally, the model may also be useful for downscaling estimates of $R_{max}$ from coarse resolution weather models \citep{Davis_2018}.

Here we have applied our model to $R_{17.5ms}$ given its numerous advantages noted above. However, our model is not specific to this choice and can be easily applied to any other desired wind radius if given sufficient data. Doing so simply requires re-estimating the model coefficients from the new dataset. Alternatively, one could also choose to first model the relationship between the new wind radius and $R_{17.5ms}$ and then apply our model as presented here. To link $R_{17.5ms}$ to a larger wind radius in the outer circulation, a model similar to that presented here could be used but with only the single predictor $\frac{1}{2}fR_{new}$ since the outer circulation structure is largely independent of $V_{max}$ \citep{Frank_1977, Merrill_1984, Chavas_Lin_2016}. Another option is to directly apply the outer wind structure component of the C15 model since it has been shown to provide an excellent representation of the hurricane outer wind structure when compared to QuikSCAT data \citep{Chavas_Lin_Emanuel_2015}.

We highlight that the quantity $\frac{M_{max}}{M_{17.5ms}}$ analyzed here represents the radially-integrated effect of surface drag on an air parcel, which may hold useful information for helping to constrain the variation of $C_d$ in the inner core of a hurricane. To our knowledge this quantity has yet to be applied in such a context.

The model as presented here was applied to two historical datasets owing to the highly variable methods for estimating wind radii, especially at the surface. One promising new dataset on the horizon is surface wind field data estimated from synthetic aperture radar (SAR) \citep{Mouche_etal_2017, Zhang_etal_2021}, which has the potential to provide a direct estimate of the near-surface wind field that comes from a single consistent source with minimal bias at all wind speeds in a hurricane. The SAR dataset is currently limited in sample size but has shown substantial promise for hurricane near-surface wind field estimation \citep{Combot_etal_2020}. As this dataset grows, applying our model to SAR data is likely to be a fruitful avenue for refining parameter estimates in future work.

%%%%%%%%%%%%%%%%%%%%%%%%%%%%%%%%%%%%%%%%%%%%%%%%%%%%%%%%%%%%%%%%%%%%%
% ACKNOWLEDGMENTS
%%%%%%%%%%%%%%%%%%%%%%%%%%%%%%%%%%%%%%%%%%%%%%%%%%%%%%%%%%%%%%%%%%%%%
\acknowledgments

We thank Galina Chirokova (CSU CIRA) for maintaining the Extended Best Track dataset and Jonathan Vigh for creating the TC-OBS dataset. DC was funded by NSF grants 1826161 and 1945113. JK thanks NOAA's Center for Satellite Applications and Research for the time, resources and funding to work on this study. The views, opinions, and findings contained in this report are those of the authors and should not be construed as an official National Oceanic and Atmospheric Administration or U.S. Government position, policy, or decision. 

%%%%%%%%%%%%%%%%%%%%%%%%%%%%%%%%%%%%%%%%%%%%%%%%%%%%%%%%%%%%%%%%%%%%%
% DATA AVAILABILITY STATEMENT
%%%%%%%%%%%%%%%%%%%%%%%%%%%%%%%%%%%%%%%%%%%%%%%%%%%%%%%%%%%%%%%%%%%%%
% 
%
\datastatement
All data analyzed in this study are freely available from the Extended Best Track website at \url{https://rammb.cira.colostate.edu/research/tropical_cyclones/tc_extended_best_track_dataset/} and the TC-OBS website at \url{https://verif.rap.ucar.edu/tcdata/historical/} .
%%%%%%

%%%%%%%%%%%%%%%%%%%%%%%%%%%%%%%%%%%%%%%%%%%%%%%%%%%%%%%%%%%%%%%%%%%%%
% APPENDIXES
%%%%%%%%%%%%%%%%%%%%%%%%%%%%%%%%%%%%%%%%%%%%%%%%%%%%%%%%%%%%%%%%%%%%%
%
% Use \appendix if there is only one appendix.
%\appendix

% Use \appendix[A], \appendix}[B], if you have multiple appendixes.
%\appendix[A]

%% Appendix title is necessary! For appendix title:

%%% Appendix section numbering (note, skip \section and begin with \subsection)
% \subsection{First primary heading}

% \subsubsection{First secondary heading}

% \paragraph{First tertiary heading}

%% Important!
%\appendcaption{<appendix letter and number>}{<caption>} 
%must be used for figures and tables in appendixes, e.g.,
%
%\begin{figure*}
%\noindent\includegraphics[width=19pc,angle=0]{figure01.pdf}\\
%\appendcaption{A1}{Caption here.}
%\end{figure*}
%
% All appendix figures/tables should be placed in order AFTER the main figures/tables, i.e., tables, appendix tables, figures, appendix figures.
%
%%%%%%%%%%%%%%%%%%%%%%%%%%%%%%%%%%%%%%%%%%%%%%%%%%%%%%%%%%%%%%%%%%%%%
% REFERENCES
%%%%%%%%%%%%%%%%%%%%%%%%%%%%%%%%%%%%%%%%%%%%%%%%%%%%%%%%%%%%%%%%%%%%%
% Make your BibTeX bibliography by using these commands:
\bibliographystyle{ametsoc2014}
\bibliography{refs_CHAVAS}
% \bibliographystyle{ametsoc2014}
% \bibliography{refs_2,refs_CHAVAS}

%%%%%%%%%%%%%%%%%%%%%%%%%%%%%%%%%%%%%%%%%%%%%%%%%%%%%%%%%%%%%%%%%%%%%
% TABLES
%%%%%%%%%%%%%%%%%%%%%%%%%%%%%%%%%%%%%%%%%%%%%%%%%%%%%%%%%%%%%%%%%%%%%
%% Enter tables at the end of the document, before figures.
%%
%

%%%%%%%%%%%%%%%%%%%%%%%%%%%%%%%%%%%%%%%%%%%%%%%%%%%%%%%%%%%%%%%%%%%%%
% FIGURES
%%%%%%%%%%%%%%%%%%%%%%%%%%%%%%%%%%%%%%%%%%%%%%%%%%%%%%%%%%%%%%%%%%%%%
%% Enter figures at the end of the document, after tables.
%%
%\begin{figure*}[h]
% \centerline{\includegraphics[width=19pc]{figure01.pdf}}
%  \caption{Enter the caption for your figure here.  Repeat as
%  necessary for each of your figures. Figure from \protect\cite{Knutti2008}.}\label{f1}
%\end{figure*}

%Figure: Conceptual diagram

%Figure: Theoretical Mmax/M17.5ms

%Figure: EBT Mfrac + Rmax predict

%Figure: TC-OBS / Aircraft-only Mfrac + Rmax predict

%Figure: Rmax predict

%Figure: EBT Mfrac + Rmax predict

%Figure: Rmax vs latitude

% Figure - scatter plots of the three methods considered in the validataion

%Figure: Bias-adjusted EBT Rmax predict

\end{document}